\documentclass[Copyrighted=False
]{ceurart}
\conference{}

\usepackage{listings}
\begin{document}

\copyrightyear{2025}
\copyrightclause{Copyright for this paper by its authors.
  Use permitted under Creative Commons License Attribution 4.0
  International (CC BY 4.0).}

\title{We're Still Doing It (All) Wrong: Recommender Systems, Fifteen Years Later}

\author[1]{Alan Said}[orcid=0000-0002-2929-0529,email=alansaid@acm.org]
\author[2]{Maria Soledad Pera}[orcid=0000-0002-2008-9204,email=m.s.pera@tudelft.nl]
\author[3]{Michael D. Ekstrand}[orcid=0000-0003-2467-0108,email=mdekstrand@drexel.edu]

\address[1]{University of Gothenburg, Gothenburg, Sweden}
\address[2]{TU Delft, Delft, Netherldans}
\address[3]{Drexel University, Philadelphia, USA}

\begin{abstract}
In 2011, Xavier Amatriain sounded the alarm: recommender systems research was ``doing it all wrong'' \cite{RecommenderSystemsWere2011}. His critique, rooted in statistical misinterpretation and methodological shortcuts, remains as relevant today as it was then. But rather than correcting course, we added new layers of sophistication on top of the same broken foundations. This paper revisits Amatriain’s diagnosis and argues that many of the conceptual, epistemological, and infrastructural failures he identified still persist, in more subtle or systemic forms. Drawing on recent work in reproducibility, evaluation methodology, environmental impact, and participatory design, we showcase how the field's accelerating complexity has outpaced its introspection. We highlight ongoing community-led initiatives that attempt to shift the paradigm, including workshops, evaluation frameworks, and calls for value-sensitive and participatory research. At the same time, we contend that meaningful change will require not only new metrics or better tooling, but a fundamental reframing of what recommender systems research is for, who it serves, and how knowledge is produced and validated. Our call is not just for technical reform, but for a recommender systems research agenda grounded in epistemic humility, human impact, and sustainable practice.
\end{abstract}


\maketitle

\begingroup
\renewcommand\thefootnote{}\footnotetext{%
  \hspace{-0.2em}\raisebox{5pt}{%
    \begin{minipage}[t]{\columnwidth}
      \footnotesize
      This is the author's version of the work. It is posted here for your personal use. Not for redistribution. The definitive Version of Record was accepted for publication in the \textit{Beyond Algorithms: Reclaiming the Interdisciplinary Roots of Recommender Systems Workshop (BEYOND 2025), September 26th, 2025, co-located with the 19th ACM Recommender Systems Conference, Prague, Czech Republic}
    \end{minipage}%
  }%
}
\endgroup

\section{Still Doing It Wrong}

In 2011, Xavier Amatriain wrote a blog post that many remember but few acted on. Ratings, like Likert scales, are ordinal. However, recommender systems routinely treat them as interval, applying Pearson correlation, computing RMSE, and optimizing linear models based on assumptions users do not make.  Around the same time, \citet{konstan2012recommender} observed that despite the community's emphasis on prediction accuracy (in light of the Netflix Prize in the mid-00s), small changes in error metrics were not the path to meaningful improvements in the user experience (so foundational to recommender systems). And yet, it cannot be denied that most of the community still today continues to chase minuscule performance gains and treats it as a win.
 Amatriain called for a ``drastic change in the way we approach these issues'' \cite{RecommenderSystemsWere2011}. 
While some of the specific technical details have changed, i.e., moving from predicting ratings to generating top-$N$ lists or considering new scenarios like session-based recommendation, the more fundamental change in assumptions never came. The field has made enormous progress in algorithmic innovation, industrial adoption, and academic visibility. Still, we have to ask: \textit{is this simply more sophisticated machinery built on the same unstable ground?}

We revisit that moment, almost fifteen years later, not out of nostalgia but necessity. Despite powerful tools, reproducibility frameworks, and expansive literature, many of the field’s underlying assumptions remain untouched or have been altered in mostly superficial ways.  The metrics change, but optimization on historical data still dominates. We argue that recommender systems research continues to prioritize optimization over understanding, evaluation over reflection, and abstraction over accountability.
We do not offer new metrics or architectures. Instead, we offer a provocation, a reflection informed by a decade and a half of scholarship, community experience, and frustrated attempts at reform. Our goal is to name persistent epistemological failures and to point toward the structural and conceptual shifts needed to address them. If recommendation is to serve people, not just datasets, then RecSys researchers and practitioners driving advances in recommender systems research must ask better questions.

\section{Same Fallacies, More Layers}

Modern recommenders rarely expose raw ratings. They operate on embeddings, graph convolutions, and attention networks. Yet many still optimize for average error or ranking metrics based on the same flawed assumptions. Behavior is treated as truth, preference as (stable) ground truth, and interaction logs as stable evidence.

Recent reproducibility analyses show that even simple algorithms, such as MostPop and ItemKNN, yield inconsistent results across frameworks due to differing defaults, metrics, and evaluation logic~\cite{akhadam2025comparative}. Even on the same framework, poor parameter tuning can lead to incorrect assumptions about which algorithms outperform others in specific contexts \cite{shehzad2023everyone}. Ranking-based methods such as BPR or LightGCN still assume that a small set of clicks reveals true preference ordering. Many models trained on implicit data are validated with metrics grounded in explicit semantics. 

Advancements in the field can sometimes bring new incorrect assumptions. For instance, sequential recommendation \citep{quadranaSequenceAwareRecommenderSystems2018} is based on the sound premise that sequence and time matter for modeling user preference and generating effective recommendations. Na\"ive sequential problem formulations like next-click prediction, however, undermine the value of that advance.
Training and evaluating a model based on its ability to predict the next item the user clicked inherently ties it to past user trajectories and the internal and external factors that influenced them. This approach prioritizes replicating historical patterns over building models that help users discover items they need \emph{more} efficiently or effectively.
This issue is compounded by the long-standing limitation of dataset-driven evaluation: models are typically evaluated on their ability to find the items the user found somehow in the past, not on their ability to \emph{improve} the user experience, e.g., identifying better items \citep{ekstrandSturgeonCoolKids2017,smuckerExtendingMovieLens32MProvide2025,canamaresShouldFollowCrowd2018}.
Off-policy evaluation \citep{jeunenRevisitingOfflineEvaluation2019,saitoEvaluatingRobustnessOffPolicy2021} seeks to adjust for the influence of the previous recommender system in product exposure, but cannot correctly evaluate recommendations of relevant items to which the user was never exposed, which is the core issue. Studies like those of \citet{chaneyHowAlgorithmicConfounding2018} quantify the impact of previous exposure policies on user experience and item exposure distribution, but do not address the fundamental evaluation problem.
Targeted data collection from users \citep{smuckerExtendingMovieLens32MProvide2025} or experts \citep{koukiLabProductionCase2020} are promising directions and grapple with the fundamental problem, but are not yet widely used and are likely difficult to scale.

The need for reform is no longer a fringe position. A 2024 Dagstuhl Seminar~\cite{bauer_et_al:DagRep.14.5.58} dedicated to recommender systems evaluation outlined major deficits in theoretical justification, fairness treatment, reproducibility practices, and long-term evaluation. Despite these community diagnoses, standard practice has changed little.

At their best, recommender systems as a field and RecSys as a community represent a broad, interdisciplinary investigation into how to effectively match users with information, entertainment, products, artists, creators, romance partners, etc., that align with their needs and interests and promote mutually-beneficial confluences of informational, artistic, personal, or commercial interests.
Early research \citep{resnickGroupLensOpenArchitecture1994, billsusHybridUserModel1999, hillRecommendingEvaluatingChoices1995} was holistic, considering data, models, user experience, and user response together in a single work.
As the field has matured, research has inevitably and necessarily become more specialized.
Maintaining balance and visibility for the range of approaches and perspectives needed for effective recommendation and recommender system evaluation is an ongoing concern and debate for the field, with the relative balance of machine learning and human-oriented research (including human-computer interaction, psychology, marketing, economics, etc.) shifting over time.
Based on the balance of topics in published papers and the venues in which different types of work appear \cite{smyth2025people}, as well as reviewer comments on submitted manuscripts, machine learning and algorithmic approaches often dominate, even to the point of reviewers objecting to human-focused or evaluative research because it does not show algorithmic improvements (the authors have multiple examples of such comments).

The result is a field that, despite the ongoing discussions, is often more concerned with recommendation as a machine learning benchmark than a human problem---or at least it appears to be the case given the vast majority of published works in this area.
Delivering meaningful, impactful recommendations that real users find useful in their lives, and assessing those recommendations, is a much bigger and much more difficult problem than improving a standard metric on a standard dataset.


\section{The Cult of Evaluation}

Most research up to this point has focused on improving the ``accuracy'' of recommender systems. We believe that this narrow focus has been misguided and even detrimental to the field. The recommendations that are most accurate according to the standard metrics are sometimes not the recommendations that are most useful to users~\cite{mcnee2006being}.
RMSE has been replaced by nDCG, Precision, and Recall, but the fundamental logic of dataset optimization remains. A systematic review of evaluation-focused recommender system research from 2017 to 2022 found that the vast majority of work uses only a few metrics and relies heavily on offline experiments~\cite{bauerExploringLandscapeRecommender2024}. Beyond-accuracy metrics such as diversity, novelty, and fairness remain rare \cite{kaminskas2016diversity}. Most papers evaluate on one or two standard datasets, often MovieLens 1M or an Amazon review subset.

Within evaluation-focused research, 32\% of papers use a single metric; over 70\% use three or fewer~\cite{bauerExploringLandscapeRecommender2024}. Many include RMSE or MAE, despite consensus that they are inadequate for assessing user-centric quality. The result is a precise but shallow evaluation culture, disconnected from actual outcomes.

Evaluation is also not the same thing as knowledge.
Reflecting on TREC\footnote{The Text REtrieval Conference (TREC) is a series of workshops that promote research in information retrieval, offering test collections, standardized evaluation methods, and a platform for result comparison \cite{voorhees1998overview}.} and its role in information retrieval research, Ian Soboroff tweeted in 2021\footnote{\url{https://web.archive.org/web/20210815134713/https://twitter.com/ian_soboroff/status/1426901262369439751}} that `The datasets were not built to be solved. They were built as tools to understand the problem and the systems we build to ``solve'' them.'

Recommender systems datasets can play a similar role: they can serve as tools to help us scientists improve our understanding of user behavior, recommendation problems, and the relative strengths and weaknesses of different approaches to recommendation.
Such knowledge is more likely to be generalizable into new problem settings and applications than improved metric performance.
To generate such knowledge, however, experiments need to be structured to produce knowledge, not just assess performance.
A simple example that is thankfully increasing in prevalence is an ablation study, that seeks to understand \textit{which} components of a proposed solution are driving observed metrics.
Similarly, disaggregated \citep{barocasDesigningDisaggregatedEvaluations2021} and distributional \citep{ekstrandDistributionallyinformedRecommenderSystem2024} evaluations aim to understand how a system's performance may vary for different types of users and items. Detailed understandings of the context of recommender system behavior and performance will also make it easier to determine whether and when proposed new advances can be applied in production.
Experimentation for knowledge needs more design like this: metrics, analyses, and runs that look beyond leaderboard benchmark performance to understand \textit{when}, \textit{why}, \textit{how}, and \textit{for who} a new idea is delivering beneficial results.

There are ongoing efforts to improve this landscape. The FEVR framework offers a structured approach to organizing recommender systems evaluation goals, metrics, and methodologies~\cite{zangerleEvaluatingRecommenderSystems2022}. The CAFE framework \cite{bauer2025conversational}, focused on conversational recommender (and search) systems,  propose evaluation methodologies that consider stakeholder goals, user tasks, user characteristics, diverse assessment criteria, methodology, and quantitative measures. Frameworks such as these aim to make evaluation more transparent, purpose-driven, and context-aware. However, such frameworks remain underutilized, and their influence on standard practice has so far been limited.

\section{Reproducibility \texorpdfstring{\boldmath$\neq$}{≠} Reliability}

Improved tooling and reproducibility tracks have increased transparency, but not comparability. A small set of datasets (MovieLens, Amazon) dominates, yet papers rarely disclose preprocessing details, timestamp filtering, or train/test leakage risks \citep{sunAreWeEvaluating2020,hidasiWidespreadFlawsOffline2023}. Nearly half of papers use a single dataset; two-thirds rely on datasets used only once~\cite{bauerExploringLandscapeRecommender2024}.

Offline evaluation continues to serve as the default setting. Even in dedicated evaluation studies, temporal dynamics are often ignored, and random splits are still widely employed \citep{mengExploringDataSplitting2020}. Empirical evidence shows that evaluation frameworks such as Elliot, RecBole, and Cornac can produce significantly different outcomes for the same algorithm, even when configurations are aligned \cite{schmidtEvaluatingPerformancedeviationItemKNN2024}. This inconsistency is not limited to complex or novel models. Performance variance has been observed even for simple baselines such as ItemKNN or MostPop.

Many experimental defaults, including relevance thresholding, sampling strategies, cutoff choices, and metric implementations, differ across frameworks and publications. These differences are rarely disclosed or justified. For example, the choice of random seed can meaningfully alter results, yet most studies report only a single run without intervals or variance estimates. In some cases, the variation caused by these defaults exceeds the claimed improvements over established baselines.

In this environment, reproducibility often amounts to rerunning fragile configurations rather than building confidence in robust insights. As long as evaluation is treated as procedural rather than interpretive, the field risks reinforcing artifacts of implementation instead of uncovering meaningful patterns. Reliability requires not only access to code and data, but also critical reflection on design decisions and their consequences.

\section{New Sins for a New Era}

As a community, we have not only failed to correct past mistakes, but we have invented new ones:
\vspace{-0.2cm}
\begin{itemize}\setlength\itemsep{2pt}
  \item \textbf{Environmental neglect}: Recommenders increasingly rely on resource-intensive architectures. Few papers disclose compute costs or carbon impact. Training one deep recommender system model can consume orders of magnitude more energy than traditional approaches \cite{venteClicksCarbonEnvironmental2024,spilloComparingDataReduction2025,spilloRecSysCarbonAtorPredicting2025}.
  \item \textbf{Unchecked need}: A significant chunk of recommender system research has embraced Large Language Models (LLMs) as a panacea, making them core to the process. This has been done without regard to the extent to which LLMs can memorize common datasets; an effect that is unsurprisingly tied to improved performance \cite{di2025llms}. Further, in some cases, these models perform on par with less resource-intensive alternatives in standard offline evaluation environments \cite{di2025llms,venteClicksCarbonEnvironmental2024}. Perhaps it is better to verify that such models are necessary and will deliver benefits commensurate with their costs before deploying them. 
  \item \textbf{Ethical fragility}: Algorithmic fairness and user autonomy are now part of the discourse, but rarely built into model or system design or evaluation. Metrics are post hoc \cite{milanoRecommenderSystemsTheir2020}.
  \item \textbf{User disempowerment}: Transparency, control, and interaction remain secondary concerns. Recommenders ``push'' rather than ``negotiate'' \cite{wangTrustworthyRecommenderSystems2024}.
  Recommendation lacks reciprocity with the system designers \citep{ekstrandBehaviorismNotEnough2016} or, in generative recommendation, with the artists, authors, and other creators of recommended work \citep{burke:post-userist}.
  \citet{belkinEthicalPoliticalImplications1976} sounded the alarm about push-oriented and sender-focused information science in 1976. Their warnings are even more apropos today as generative modalities converge with disinformation and influence campaigns within the recommender-mediated platforms that shape much of how people understand the world.
\end{itemize}

\section{What Would Doing It Right Look Like?}

Doing it right does not mean optimizing harder. It means asking better questions and reflecting on the answers. There is no need for a single metric or framework; reducing improvements to standardized procedures is more likely to hinder the necessary diversity of research methods and lines of inquiry needed to keep recommender systems on a course of human welfare. However, recommender systems research needs: (i) Diverse datasets, evaluated across varied contexts; (ii) Human-aligned evaluation: not only offline precision, but meaningful outcomes; (iii) Transparent reporting: preprocessing, hyperparameters, compute, and code; (iv) Sustainable practices: energy as a first-class consideration; (v) Epistemic humility: acknowledging noise, preference volatility, and modeling limits; and (vi)  Normative and human grounding: explicit articulation of the individual or human social goals of the system or evaluation, and connection of the technical work to those goals.


Recent work has emphasized the societal obligations of recommender system researchers. In the context of RecSys for Social Good, it has been argued that recommender systems research must be responsive not only to users' preferences, but also to broader social outcomes such as justice, inclusion, well-being, and sustainability development goals~\cite{saidRecommenderSystemsSocial2024,jannachRecommenderSystemsGood2025,felfernig2023recommender,patro2020towards}. Doing it right, then, requires not just methodological reform, but a reframing of what we consider valuable progress in the field.

Community-led initiatives have emerged to create space for alternative voices and perspectives. The AltRecSys workshop (at RecSys 2024) explicitly foregrounded speculative, critical, and unconventional ideas in recommendation~\cite{ekstrandAltRecSysWorkshopAlternative2024}. Rather than showcasing polished technical contributions, the workshop invited participants to question foundational assumptions, share negative results, and propose revisiting what counts as valuable research. This venue demonstrated that the community has become increasingly aware that many of the field's core challenges are epistemological and institutional as much as they are technical. The NORMalize  \cite{vrijenhoek2023normalize} and RecSoGood \cite{boratto2024first} workshops (co-located with RecSys) encouraged the community to rethink evaluation practices, highlighting the importance of considering norms and values driving recommender systems, along with their potential to guide users towards more sustainable choices and behaviors, supporting broader environmental and social goals. 
At the risk of tooting our own horn, ongoing work on the LensKit Codex\footnote{\url{https://codex.lenskit.org}} aims to improve consistency and sustainability of baseline comparisons by providing standardized evaluation results and hyperparameter tunings for common models on a range of public datasets, allowing researchers to compare against well-tuned baseline models without incurring the environmental cost (and risk of error) of re-tuning the baselines themselves, and to invite community collaboration on tuning methods and standardized results. 

Recent work has proposed reimagining recommender system development as a participatory process, where users, providers, and other stakeholders act not only as research subjects, but as co-designers and co-evaluators of the systems that shape their digital experiences~\cite{ekstrand2025co}. This approach challenges the prevailing notion that designers alone should define the goals and metrics. Instead, it advocates for a redistribution of design authority, grounded in democratic values and the lived experiences of those affected. Doing it right, in this view, means sharing power, not just optimizing performance. 
\citet{burke:post-userist} argue for recommender systems research to make a \textit{relational} turn (or re-turn, as key early research had a decidedly relational bent \citep{hillRecommendingEvaluatingChoices1995}).
This sentiment is also prevalent in recent works focused on multistakeholder-focused evaluation \cite{BURKE2025103560}, arguing to move beyond accuracy measures that simply capture the overall utility of a single stakeholder, to consider the complex, domain-dependent, and multifaceted task of probing the impact of recommender systems on all of their stakeholders.

Holistic, human-centered recommender systems research is not \textit{easy}.  It often takes more time to plan and execute, and may require resources beyond those needed by algorithmic experiments on published data.
Several projects are working to make grounded, in-vivo evaluation more accessible, such as Informfully \citep{heitzInformfullyResearchPlatform2024} and POPROX\footnote{\url{https://poprox.ai}} \citep[co-developed by the last author]{burkeConductingRecommenderSystems2025}, as well as the earlier CLEF NewsReel challenge organized with Plista \citep{brodtSheddingLightLiving2014}. Some RecSys Challenges, like the one in 2024 \cite{recsyschallenge24}, focused not only on accuracy but also on beyond-accuracy metrics to account for the normative complexities inherent in news recommendations; the 2021 one \cite{recsyschallenge21} aimed for multi-goal optimization, prompting participants to predict user engagement probabilities for tweets while ensuring recommendations were both accurate and fair. 
These platforms and challenges provide infrastructure that can support more holistic research, as well as working examples for the development of further resources. 
However, the community has yet to consistently adopt them, let alone sustain their use in the long term.

Echoing the argument of \citet{olteanuRigorAIDoing2025}, in the end, rigorous research on recommender systems requires going beyond technical and statistical notions of rigor to include normative, epistemic, and other forms of rigor as well.
Encouraging the community towards a well-rounded evaluation framework, one that considers multiple angles to probe a recommender system and convey if it is `good' (in its architecture, performance, but also for whom, from which perspective, and in which context), is already supported in the literature \citep[e.g.,][]{bauer2024values,stray2024building,rampisela2024can,ungruh2025mirror}. We ``just'' need to embrace it and take it from theory to practice, and make closing those gaps a priority for the field and community.  Table~\ref{tab:paradigm-shift} summarizes several persistent structural defaults in recommender systems research and highlights what a more human-centered, reflective, and accountable paradigm might look like. These contrasts are not exhaustive, but they capture the epistemic shift we argue is necessary: from narrow optimization toward broader alignment with user and societal values.

\begin{table}[h]
\centering
\caption{Shifting recommender systems practice}
\label{tab:paradigm-shift}
\resizebox{.7\textwidth}{!}{
\begin{tabular}{lll}
\toprule
\textbf{Dimension} & \textbf{Status Quo} & \textbf{Better Alternative} \\
\midrule
\makecell[l]{Assumptions \\about data} & \makecell[l]{Preferences are stable;\\ ordinal = interval} & Preferences are noisy, contextual \\
\midrule
Evaluation focus & RMSE, nDCG, Precision & Human goals, mixed methods \\
\midrule
System goal & Maximize clicks or accuracy & Support reflection, fairness, trust \\
\midrule
Model validation & One-off benchmarks & Transparent, replicable, contextual \\
\midrule
User role & Target of optimization & Stakeholder, co-designer \\
\bottomrule
\end{tabular}
}
\vspace{-0.5cm}
\end{table}

\section{Conclusion: Enough Already}

Reflecting in 2004 about evaluation of collaborative filtering recommender systems, \citet{herlocker2004evaluating} wrote that ``Effective and meaningful evaluation of recommender systems is challenging. To date, there has been no published attempt to synthesize what is known [...], nor to systematically understand the implications [...] 
for different tasks and different contexts.'' Today, there are more surveys and frameworks, but a thorough and systematic assessment for collaborative filtering or the many model families extant in the field does not seem much closer now than it did 20 years ago.

Amatriain's critique remains true. Recommender systems research is still doing it wrong. It is doing so at scale, with better tools, more complex models, and increased compute.
This is not a problem of ignorance. The challenges are well-documented: from the ordinal nature of ratings or the inadequacy of top-k metrics and implicit feedback as proxies for user experience and ground truth, to the costs of large models. Researchers, developers, and reviewers are aware of these issues. Yet the publication and reward structures in the field continue to favor narrow performance gains and apparent (often mathematical) sophistication over conceptual clarity, ecological awareness, or social relevance \citep{liptonTroublingTrendsMachine2019}.

It is time to stop pretending that stronger baselines, larger datasets, new benchmarks, or even more rigorous experimental protocols will fix foundational misalignments.  Recommender systems research must shift from engineering models to understanding influence and impact.

Fixing this requires recommender systems research to adopt an entirely different posture, one of epistemic humility. We must acknowledge the limitations of our methods and data, and treat evaluation not as a procedural step, but as a site of inquiry. We must ask what kinds of knowledge recommender systems research should produce, and for whom.
The point is not to stop working on models and algorithms, but to do better and more useful research on models and algorithms: to contextualize and evaluate them in terms of how well they --- and the research and engineering methods behind them --- actually drive meaningful improvements in the human experience of recommender systems. There are many gaps between current practice, capabilities, and knowledge and the kinds of human impact the field looks to have, providing important problems and settings for future research.

Recommender systems are not just algorithms; they are sociotechnical interventions \cite{seaverAlgorithmsCultureTactics2017,stray2024building,lucheriniTRECSSimulationTool2021,mitraSearchSocietyReimagining2024}. They shape what people see, believe, and desire. If we do not take this responsibility seriously, we will continue to develop systems that are efficient and elegant, but ultimately misaligned with the people they are meant to serve.

Let us stop solving the wrong problem better. Let us define a better problem.
\section*{Declaration on Generative AI}
During the preparation of this work, the author(s) used X-GPT-4 and Grammarly for Grammar and
spelling check and rephrasing. After using these tool(s)/service(s), the author(s) reviewed and edited the content as needed and take(s) full responsibility for the publication’s content.
\bibliography{sample}

\end{document}